\documentclass[11pt,a4paper]{article}

\usepackage[T1]{fontenc}
\usepackage[utf8]{inputenc}
\usepackage{lmodern}
\usepackage{amsmath,amssymb,amsthm}
\usepackage{geometry}
\usepackage{graphicx}
\usepackage{hyperref}
\usepackage{booktabs}
\usepackage{enumitem}
\usepackage{setspace}
\usepackage{array}
\usepackage{microtype}
\usepackage{caption}
\usepackage{float}
\usepackage{longtable}

\hypersetup{
    colorlinks=true,
    linkcolor=blue,
    citecolor=blue,
    urlcolor=blue,
    pdftitle={ECCFROG522PP: An Enhanced 522 bit Weierstrass Elliptic Curve},
    pdfauthor={Victor Duarte Melo}
}

\title{\textbf{ECCFROG522PP: An Enhanced 522 bit Weierstrass Elliptic Curve}}

\author{
V\'ictor Duarte Melo\\
Independent Researcher\\
Erd\H{o}s Number: 4
}

\date{}

\begin{document}

\maketitle

\begin{abstract}
This paper presents ECCFROG522PP, a 522 bit elliptic curve over a prime field in short Weierstrass form, designed around a simple principle: every critical parameter should be publicly reproducible from a fixed seed by a deterministic procedure. Many deployed systems still rely on NIST P 256 and secp256k1, which sit near the 128 bit classical security level. At higher security levels, practitioners usually consider NIST P 521, Curve448, and Brainpool P512. ECCFROG522PP is intended for the same general classical security range as P 521 while emphasizing transparency, verifiability, and auditability rather than speed. The curve parameters are derived from a fixed public seed through a BLAKE3 based pipeline with published indices. The resulting curve has prime order, cofactor one, a deterministically validated base point, a quadratic twist with a large proven prime factor, a published embedding degree lower bound, and basic sanity checks against small embedding degree reductions and low bound CM anomalies. The complete generation procedure can be reproduced and checked end to end from public artifacts and implementation scripts.
\end{abstract}

\section{Introduction}

Elliptic curve cryptography was introduced independently by Koblitz and Miller in the middle of the 1980s \cite{Koblitz1987,Miller1985}. The subject later matured into one of the most widely deployed public key technologies in modern systems. The practical appeal of elliptic curves comes from the fact that one can obtain strong security with key sizes that are much smaller than those of many classical alternatives. Over time, a small number of curves came to dominate deployments, notably secp256k1 in blockchain ecosystems and NIST P 256 in many software and protocol stacks.

At the same time, the question of how curve parameters are generated has remained important. In some designs, parameter generation is easy to explain and reproduce. In others, historical documentation or design rationale leaves room for discomfort among implementers and reviewers. This concern is not merely aesthetic. In cryptography, transparency of construction contributes to confidence. If a curve can be regenerated by an independent party from a small set of public rules, then trust can shift away from the authority that first published it and toward a process that anyone can verify.

ECCFROG522PP was designed with that goal in mind. It is a prime field elliptic curve in short Weierstrass form whose core parameters are derived from a fixed public seed through deterministic BLAKE3 based rules. The design deliberately minimizes discretionary choices. The field modulus is fixed. The coefficient generation is fixed. The base point selection rule is fixed. The search indices are published. The intended outcome is a curve whose origin is easy to inspect and whose validation can be automated by script.

The curve is also integrated into the open source HippoFrog toolchain, which uses ECCFROG522PP in a file encryption setting. The implementation context is useful because it shows that a transparency focused curve need not remain purely theoretical. However, the mathematical definition of the curve stands independently of any one software package. The HippoFrog repository is available at
\[
\texttt{https://github.com/victormeloasm/HippoFrog}.
\]

The contribution of this paper is therefore straightforward. It documents a reproducible 522 bit curve, gives its main parameters, explains the derivation rules, summarizes the published sanity checks, and frames the design as a practical transparency driven alternative in the same broad classical security range usually associated with NIST P 521.

\section{Background}

A short summary of the main families of elliptic curves used in practice is helpful for context.

An Edwards curve can be written as
\begin{equation}
x^2 + y^2 = 1 + d x^2 y^2 \pmod p.
\end{equation}
Ed25519 is the most familiar example.

A Montgomery curve can be written as
\begin{equation}
B y^2 = x^3 + A x^2 + x \pmod p.
\end{equation}

A short Weierstrass curve has the form
\begin{equation}
y^2 = x^3 + ax + b \pmod p.
\end{equation}

ECCFROG522PP belongs to the short Weierstrass family. In that model, a curve is specified by the field prime \(p\), coefficients \(a\) and \(b\), a base point \(G = (G_x, G_y)\), and the order \(N\) of the subgroup generated by \(G\). Standard validation concerns include non singularity, the subgroup structure, resistance against invalid curve failures under proper validation, and the absence of unexpectedly small embedding degree phenomena.

This paper does not claim a new family of elliptic curves. Instead, it focuses on a different design emphasis inside a familiar family. The main novelty is not algebraic form. It is deterministic generation, explicit publication of the derivation path, and the attempt to reduce ambiguity about origin.

\section{Motivation and Design Philosophy}

Transparent generation of cryptographic parameters is attractive for at least three reasons.

First, it improves auditability. An independent reviewer should be able to inspect not only the final published constants but also the process that produced them. If that process is deterministic and fixed in advance, the amount of trust placed in the original publisher is reduced.

Second, it supports long term reproducibility. A curve that depends on undocumented historical choices or on unavailable intermediate material is harder to evaluate decades later. By contrast, a curve derived from a fixed public seed and a small set of explicit rules can be rechecked whenever new tools or new concerns emerge.

Third, it creates a cleaner narrative for adoption in experimental systems. Even if a new curve is not intended to replace a standard immediately, a reproducible construction can still be valuable as a research artifact, a benchmark target, or a tool integration point.

ECCFROG522PP was therefore designed under the following principles.

\begin{enumerate}[label=\arabic*.]
    \item Public and deterministic derivation of core parameters.
    \item Prime order subgroup with cofactor one.
    \item No speed claim as a design objective.
    \item Straightforward independent validation by script.
    \item Practical integration in a real software artifact.
\end{enumerate}

The curve should be read in that spirit. It is not presented as a universal replacement for all established curves. It is presented as a transparency focused construction in the P 521 class security range, supported by explicit derivation rules and public artifacts.

\section{Curve Definition}

ECCFROG522PP is defined over a prime field \(\mathbb{F}_p\) and uses the short Weierstrass equation
\begin{equation}
y^2 = x^3 - 9x + b \pmod p.
\end{equation}

The public seed used throughout the construction is
\[
\texttt{ECCFrog522PP|v1}.
\]

The published search indices are

\begin{itemize}
    \item coefficient index \(i = 1{,}294{,}798\),
    \item base point index \(j = 0\).
\end{itemize}

The fact that \(j = 0\) already yields a full order base point is simply the observed output of the deterministic procedure.

\subsection{Field modulus}

The field prime is

\begin{flushleft}\ttfamily\small
686479766013060971498190079908139321726943530014330540939446345918554318\\
339765605212255964066145455497729631139148085803712198799971664381257402\\
8291115058039
\end{flushleft}

\subsection{Curve coefficient}

The deterministic coefficient \(b\) is

\begin{flushleft}\ttfamily\small
661139136184195850860452469937744791138999490012975421307768311225096419\\
509388251093415492337101182055425457255989613682399356563300695566619742\\
8760619911
\end{flushleft}

\subsection{Group order}

The subgroup order is

\begin{flushleft}\ttfamily\small
686479766013060971498190079908139321726943530014330540939446345918554318\\
339765470783993099806907243717889863432321841973824511791072608043490749\\
5541251156283
\end{flushleft}

This order is prime, so the cofactor equals one.

\subsection{Base point}

The deterministic base point coordinates are

\begin{flushleft}\ttfamily\small
G\(_x\) = 114836598700559139646235363713136312609767670986199491984058026550790121\\
317888159000151000981405923011587990724012666535482931446873066751491073\\
89798128134
\end{flushleft}

\begin{flushleft}\ttfamily\small
G\(_y\) = 303869445742844202438813211737067794312734393851211346303431863870960045\\
113632574702513861080239149191409127648110569935391920249490281068659303\\
0172286395020
\end{flushleft}

The point \(G\) has full order \(N\).

\section{Deterministic Generation Procedure}

The central feature of ECCFROG522PP is that its key parameters are generated through a fixed, public, deterministic procedure.

\subsection{Coefficient derivation}

The coefficient \(b\) is derived by the rule
\begin{equation}
b = \left( \mathrm{BLAKE3}(\texttt{seed}\,\|\,\texttt{b}\,\|\,i) \bmod (p - 3) \right) + 2.
\end{equation}

This constrains \(b\) to the interval \([2, p - 2]\), which avoids trivial and degenerate cases. The search process then checks the resulting candidate against the required curve criteria and retains the first acceptable output at the published index.

\subsection{Base point derivation}

The \(x\) coordinate of the candidate base point is generated by
\begin{equation}
G_x = \mathrm{BLAKE3}(\texttt{seed}\,\|\,\texttt{G}\,\|\,j) \bmod p.
\end{equation}

Given this candidate, the corresponding point on the curve is recovered and then validated. In the published construction, the first checked index already yields a full order point.

\subsection{Digest size and determinism}

The derivation uses BLAKE3 with a 64 byte digest for both \(b\) and \(G_x\) candidates, followed by reduction modulo the appropriate field quantity. Because the seed, constants, and indices are public, the full construction can be regenerated exactly.

\section{Security Checks and Structural Sanity}

The published checks are intentionally conventional. Their purpose is not to prove absolute security, which no finite set of practical tests can do. Rather, the goal is to rule out several common and undesirable structural weaknesses.

\subsection{Prime order and cofactor}

The order \(N\) is prime, and therefore the cofactor is one. This is a desirable property in practical protocols because it simplifies subgroup handling and removes entire classes of small subgroup concerns inside the main group.

\subsection{Frobenius trace and CM discriminant}

The Frobenius trace is

\begin{flushleft}\ttfamily\small
134428262864259238211779839767706826243829887687008899056337766653274986\\
3901757
\end{flushleft}

This yields the discriminant like quantity
\begin{equation}
D = t^2 - 4p.
\end{equation}

The corresponding published decimal value is

\begin{flushleft}\ttfamily\small
-25652094854852200923182489755562709400813783410907538423881259128294063\\
827498368666061775444493529979367517268977200639940503231230605133844631\\
506932712545107
\end{flushleft}

The reported CM sanity check passes small square free tests up to 100k.

\subsection{Embedding degree checks}

The anti MOV style sanity procedure reports no reduction of the form
\[
p^k \equiv 1 \pmod N
\]
for \(k \leq 200\). The corresponding published lower bound on the embedding degree is 14.

\subsection{Quadratic twist}

The quadratic twist order is

\begin{flushleft}\ttfamily\small
686479766013060971498190079908139321726943530014330540939446345918554318\\
339765739640518828325383667277569398845974329633599885808870720719024056\\
1040978959797
\end{flushleft}

A largest proven prime factor of approximately 505 bits divides this order. Under ordinary validation assumptions, this provides the expected level of confidence against invalid curve style failures that exploit weak twist structure.

\section{Reproducibility Workflow}

A curve intended for transparency should be easy to verify independently. The intended workflow for ECCFROG522PP is as follows.

\begin{enumerate}[label=\arabic*.]
    \item Start from the public seed \texttt{ECCFrog522PP|v1}.
    \item Recompute the coefficient \(b\) from the canonical BLAKE3 rule and the published coefficient index.
    \item Construct the curve over \(\mathbb{F}_p\) with \(a = -9\).
    \item Recompute the deterministic base point candidate from the published base point index.
    \item Verify that the resulting point has order \(N\).
    \item Recompute or independently confirm the group order.
    \item Re run the anti MOV bound check, the CM sanity checks, and the twist factor verification.
\end{enumerate}

The reference implementation of this workflow is given as a SageMath based script with BLAKE3 integration and machine readable reporting output. This procedure is the normative way to validate the publication.

\subsection{Original search environment}

The original search environment was reported as Linux 6.14 on a Ryzen 9 5950X with 32 logical cores, roughly 130 GB of memory, and a runtime near 216k seconds, or about 60 hours. These details are useful for replication context, but they are not required for independent verification of the final published parameters.

\section{Practical Integration in HippoFrog}

ECCFROG522PP is integrated into the HippoFrog file encryption tool. In the implementation summary associated with the curve, the operational pipeline uses ephemeral ECDH, HKDF SHA 256, and AES 256 GCM, together with a deterministic file header and a parameter hash binding in the header. Those engineering details are separate from the mathematical curve definition, but they demonstrate that the curve can be placed inside a working software system.

This distinction matters. A curve paper should not blur the line between abstract parameter design and implementation practice. At the same time, software integration is useful evidence that the construction is concrete enough to be tested, benchmarked, and exercised by real code.

For the present work, HippoFrog serves that role. It gives a practical home for the curve while leaving the mathematical specification independent and fully inspectable.

\section{Benchmarking Context}

The benchmark harness associated with the published material compares scalar multiplication and ECDH throughput across several standard curves and ECCFROG522PP. The point of this figure is not to claim that ECCFROG522PP is faster than P 521. The point is to position the curve empirically and honestly.

The design objective of ECCFROG522PP is transparency and reproducibility. A transparency focused curve can still be benchmarked, and such measurements are useful for implementers, but they do not alter the central claim of the work. Users should assume performance comparable to the general size class of the curve and should perform controlled measurements in their own environment if throughput is relevant to deployment decisions.

\begin{figure}[H]
    \centering
    \includegraphics[width=\textwidth]{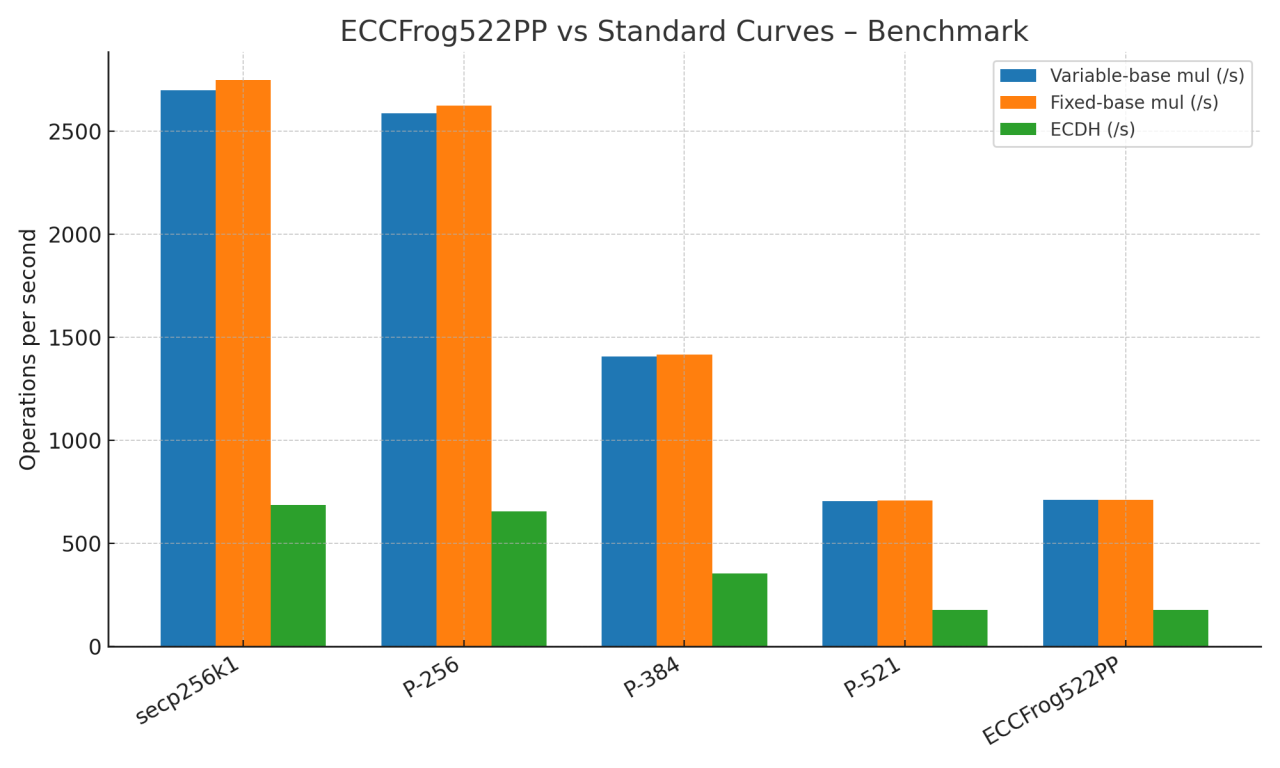}
    \caption{Benchmark comparison between ECCFROG522PP and common standard curves.}
    \label{fig:benchmark}
\end{figure}

Figure \ref{fig:benchmark} shows the benchmark plot included with the saved publication materials. The figure compares variable base scalar multiplication, fixed base scalar multiplication, and ECDH throughput across secp256k1, P 256, P 384, P 521, and ECCFROG522PP. The intended interpretation is modest. The figure illustrates practical positioning. It is not a claim of universal superiority, and it should be understood together with implementation details and measurement conditions.

\section{Positioning Relative to Existing Curves}

A sensible way to interpret ECCFROG522PP is as a curve with conventional mathematical form and non conventional design emphasis. It does not attempt to displace Edwards curves in software that already benefits from their arithmetic properties. It does not attempt to replace highly tuned standard curves on performance grounds. Instead, it offers a different value proposition.

Compared with NIST P 256 and secp256k1, the main distinction is security level and transparency framing. Compared with P 521, the most relevant distinction is that ECCFROG522PP makes deterministic generation and fully public derivation central to the narrative. Compared with Brainpool style thinking, it similarly emphasizes explicit reproducibility.

This means the curve is best suited for contexts where parameter origin and auditability are part of the selection criteria. Research prototypes, experimental cryptographic software, comparative studies, and trust focused toolchains are natural examples.

\section{Limitations}

The following limitations should be stated clearly.

\begin{enumerate}[label=\arabic*.]
    \item The curve is not presented as faster than P 521.
    \item The published anti MOV and CM related checks are bounded exactly as stated and can be extended by independent reviewers.
    \item The work emphasizes deterministic origin and reproducibility, not a new arithmetic family or a new speed record.
    \item As with any non standard curve, broad adoption would require extensive independent review and implementation scrutiny.
\end{enumerate}

These limitations do not weaken the main contribution. They clarify it.

\section{Conclusion}

ECCFROG522PP shows that a practical 522 bit prime field curve in short Weierstrass form can be presented with a fully inspectable and reproducible origin. Every important construction step begins from a public seed and follows a fixed deterministic rule. The coefficient path is explicit. The base point path is explicit. The indices are explicit. The subgroup order, twist information, and sanity checks can be re verified independently.

In that sense, the curve contributes a transparency first perspective to the familiar P 521 security class. Its main value is not speed, nor a claim of revolutionary mathematics, but rather the reduction of trust in unexplained historical choices. The project invites verification instead of asking for confidence by authority.

\appendix

\section{Selected Published Facts}

\begin{longtable}{>{\raggedright\arraybackslash}p{0.27\textwidth}p{0.67\textwidth}}
\toprule
Item & Value \\
\midrule
Seed & \texttt{ECCFrog522PP|v1} \\
Coefficient index & \(1{,}294{,}798\) \\
Base point index & \(0\) \\
Field model & Prime field short Weierstrass curve \\
Equation & \(y^2 = x^3 - 9x + b \pmod p\) \\
Field size & 522 bit prime field \\
Subgroup order & 521 bit prime \\
Cofactor & \(1\) \\
Embedding degree note & Lower bound 14 \\
Anti MOV check & No hit for \(k \leq 200\) \\
Twist evidence & Large proven prime factor of about 505 bits \\
Implementation context & HippoFrog, MIT licensed \\
\bottomrule
\end{longtable}

\section{Validation Guidance}

An independent validator may use the following checklist.

\begin{enumerate}[label=\arabic*.]
    \item Confirm that the prime \(p\) matches the published modulus.
    \item Derive \(b\) from the public seed and coefficient index.
    \item Verify that the resulting curve is non singular.
    \item Regenerate the base point candidate from the base point index.
    \item Confirm that the point lies on the curve.
    \item Confirm that the point has order \(N\).
    \item Recompute the trace and discriminant related value.
    \item Re run the anti MOV search bound and twist factor checks.
    \item Compare all outputs against the published artifact set.
\end{enumerate}

\section{Artifact Note}

The article level artifacts associated with this paper are intended to include the curve specification, the reproduction script, machine readable outputs, and benchmark material. The software integration target is HippoFrog, whose repository URL is included in the main body of the paper. These artifacts support reproducibility and make it easier for third parties to audit both the mathematical constants and the practical implementation context.


\begin{thebibliography}{99}

\bibitem{Koblitz1987}
N. Koblitz,
\newblock ``Elliptic curve cryptosystems,''
\newblock \emph{Mathematics of Computation}, vol. 48, no. 177, pp. 203--209, 1987.

\bibitem{Miller1985}
V. S. Miller,
\newblock ``Use of elliptic curves in cryptography,''
\newblock in \emph{Conference on the Theory and Application of Cryptographic Techniques},
Springer, 1985, pp. 417--426.

\bibitem{Lenstra1987}
H. W. Lenstra Jr,
\newblock ``Factoring integers with elliptic curves,''
\newblock \emph{Annals of Mathematics}, pp. 649--673, 1987.

\bibitem{BernsteinSafeCurves}
D. J. Bernstein,
\newblock ``SafeCurves: Introduction,''
\newblock \url{https://safecurves.cr.yp.to/},
accessed 04-09-2025.

\bibitem{DiSSECTNIST}
DiSSECT,
\newblock ``DiSSECT,''
\newblock \url{https://dissect.crocs.fi.muni.cz/standards/nist},
accessed 04-09-2025.

\end{thebibliography}
\end{document}